\documentclass[aps,prl,twocolumn,groupedaddress]{revtex4}%
\usepackage{graphicx}
\usepackage{amsmath}
\usepackage{subfigure}
\usepackage{siunitx}
\usepackage{color}
\usepackage[colorlinks = true, linkcolor = darkblue, urlcolor  = darkblue,
citecolor = darkblue, anchorcolor = darkblue]{hyperref}%
\usepackage{amsfonts}%
\usepackage{amssymb}
\definecolor{darkblue}{rgb}{0.0,0.0,0.6}

\begin{document}

\title{Pinning Transition of Bose-Einstein Condensates in Optical Ring Resonators}
\author{S. C. Schuster} \email[]{simon.schuster@uni-tuebingen.de}
\author{P. Wolf} 
\author{D. Schmidt} 
\author{S. Slama} 
\author{C. Zimmermann} \email[]{claus.zimmermann@uni-tuebingen.de}
\affiliation{Physikalisches Institut, Eberhard-Karls-Universit\"{a}t T\"{u}bingen, Auf der
Morgenstelle 14, D-72076 T\"{u}bingen, Germany.}

\begin{abstract}
We experimentally investigate the dynamic instability of Bose-Einstein
condensates in an optical ring resonator that is asymmetrically pumped in both
directions. We find that, beyond a critical resonator-pump detuning, the system
becomes stable regardless of the pump strength. Phase diagrams and quenching
curves are presented and described by numerical simulations. We discuss a
physical explanation based on a geometric interpretation of the underlying
nonlinear equations of motion.

\end{abstract}
\maketitle

For several years, atomic quantum gases in optical resonators have been
successfully used to study basic many-body physics with long-range
interaction. Quantum phase transitions, supersolid phases and the realization
of synthetic gauge fields are some of the current topics of this field 
\cite{Hemmerich15a, Landig, Leonard17a, Lev, Stamper-Kurn, Ritsch18}. 
While, so far, most experiments have been performed with standing wave resonators, the specific properties of ring
resonators are now coming to the fore again \cite{Goldwin18, Bouyer}. In contrast to
standing wave resonators, in an ideal ring resonator, the position of the nodes
and antinodes of an optical standing wave is not determined by end mirrors.
However, in the presence of atoms and with sufficiently strong pumping power,
this continuous symmetry can be broken spontaneously. The associated
instability was already predicted in 1998 and interpreted as an analogy to the
free-electron laser \cite{Piovella01}. The effect was also observed
experimentally more than a decade ago \cite{Kruse03, Slama07}, but, only
recently, was it possible to record a complete stability diagram \cite{Dag14, Tomczyk15}. 
These latest experiments also confirmed a model that
interprets the instability as a generalization of the Dicke phase transition
\cite{Emary03b, Zurich, Hemmerich15a}. Experiments with ring
resonators pumped simultaneously in both directions have not yet been
conducted. This Letter makes a first contribution in this direction. 

\begin{figure}[t]
\includegraphics[width=8.0cm]{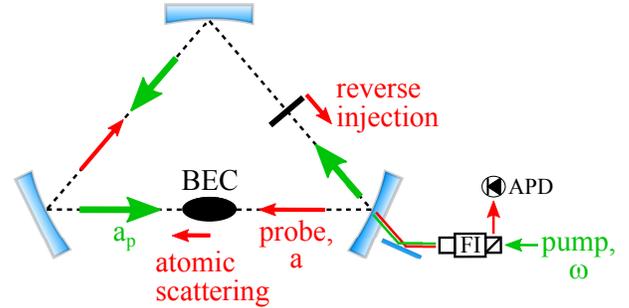}\caption{Experimental setup of a BEC 
placed in a high finesse TEM$_{00}$ mode of a ring
resonator. Light from the pump mode (green, $s$ pol.) is scattered into the
probe mode (red) by the atoms and by coherent scattering at the mirror (here,
represented by an effective reflecting element labeled "reverse injection").
The power in the probe mode is monitored by recording the probe light that
leaves the cavity through the input coupling mirror. It is separated from the
pump light with a Faraday isolator and detected with an avalanche photodiode
(APD). }%
\label{setup}%
\end{figure}

In a longitudinally pumped ring cavity, as shown in Fig. \ref{setup}, the
prominent effect is an exponential instability that is observed above a
critical pump power: If some light is present in the probe mode, the
interference pattern between the pump and the probe light generates a periodic
optical potential, which structures the initially flat atomic density
distribution. The resulting density grating efficiently diffracts pump light
into the probe mode. This, in turn, deepens the optical lattice, and also, as a
consequence, the atomic density grating increases its contrast and so on.
During the process, momentum is constantly transferred from the pump mode to
the probe mode and the atoms accelerate into the direction of the pump light
(to the right in Fig. \ref{setup}). Parallel to the atomic motion, the Doppler
effect shifts the frequency of the diffracted probe light to lower
frequencies. In this work, we extend the scenario and inject some light into
the probe mode that has the same frequency as the pump light. Together with
the pump light, it forms a stationary optical lattice that might force the
atoms to rest and suppresses the instability. Surprisingly, we find that there
is a critical detuning of the cavity relative to the pump light. Above this
detuning, the system is always stable. Below the critical detuning, the system is
still unstable for large enough pump power. In this Letter we experimentally
investigate this yet unknown ``pinning transition" and compare our observations
with numerical simulations of the nonlinear equations of motion. Furthermore,
we present a geometric interpretation of the equations, which reveals the
underlying physical mechanism.

The experimental setup in Fig. \ref{setup} is similar as described in
\cite{Dag14}; however, now, use a much larger resonator with a round trip
length of \SI{39}{\centi\meter}, a beam waist at the position of the
condensate of $w_{0}=$ \SI{170}{\micro\meter} and a mode volume of $V=$
\SI{18.2}{\milli\meter}$^{3}$. For $s$-polarized light, the decay rate for the
electric field amplitude in the resonator amounts to $\kappa=2\pi\cdot$\SI
{5}{\kilo\hertz} which is about three times smaller than the recoil shift
$\omega_{r}=2\hbar k^{2}/m=2\pi\times$\SI{14.5}{\kilo\hertz} due to momentum
absorption of an initially nonmoving atom that scatters a photon from a pump
beam into the probe beam. Here, $k$ and $m$ are the wave vector of the pump
light and the mass of the atom. For $p$-polarized light, we observe a three times
larger decay rate. The $s$-polarized forward propagating TEM$_{00}$ mode (``pump
mode") is longitudinally pumped from one side with up to \SI{6}{\milli\watt}
from an amplified diode laser system at a frequency $\omega$ detuned by
$\Delta_{a}=\omega-\omega_{0}=$\SI{-60}{\giga\hertz} relative to the atomic
transition frequency $\omega_{0}$ ($D$1 Line: $5s_{1/2},F=2$ to $5p_{1/2},F=2$).
Part of the laser output is used to electronically stabilize the laser to the
reverse propagating $p$-polarized TEM$_{10}$ mode \cite{Bux07} with a precision
of about $2\pi\times$\SI{200}{\hertz}. Frequency and amplitude of the pump
light is controlled by an acousto-optical modulator. The pump frequency
$\omega$ can be tuned relative to the resonance frequency $\omega_{c}$ of the
TEM$_{00}$ mode over a range of $\Delta_{c}=\omega_{c}-\omega=\pm10\omega_{r}%
$. The power in the TEM$_{00}$ pump mode and in the counterpropagating
TEM$_{00}$-mode (probe mode) is monitored by recording the light leaking out
of the resonator mirrors with sensitive avalanche diodes.

The pinning potential required to suppress the instability beyond the critical
detuning is very small such that we don't have to inject the probe mode
externally, but rather exploit coherent scattering of pump light into the
probe mode due to inhomogeneities in the mirror coatings. The scattered light
from the three mirrors interferes according to their relative positions and to
the wavelength of the light \cite{Krenz07}. The total mirror scattering can be
varied up to a factor of 3 by controlling the position of one of the
mirrors with a piezoelement. In the experiment, the total mirror scattering
rate $\kappa_{s}$ $=\kappa\sqrt{\varepsilon}$ is determined for each
cycle by recording the resonant power ratio $\varepsilon$ of the pump and the
probe mode right before the atoms are loaded into the cavity. The magnetically
trapped Bose-Einstein condensate of $^{87}$Rb atoms is placed at the intensity
minimum in the center of the TEM$_{10}$ mode where the atoms are least
affected by the locking light. During preparation of the condensate, the laser
beams are switched off and held at one fixed frequency for about \SI
{20}{\second}. Once the condensate is in place, the locking is reactivated
within \SI{300}{\milli\second} and the pump light is then ramped up to a final
value within \SI{50}{\micro\second}, slow enough to avoid ringing of the high
finesse resonator. After a holding time of \SI{1.5}{\milli\second}, the atoms
are released from the trap and the population of the momentum states are
derived from absorption images after \SI{35}{\milli\second} of ballistic
expansion. \begin{figure}[t]
\includegraphics[width=9cm]{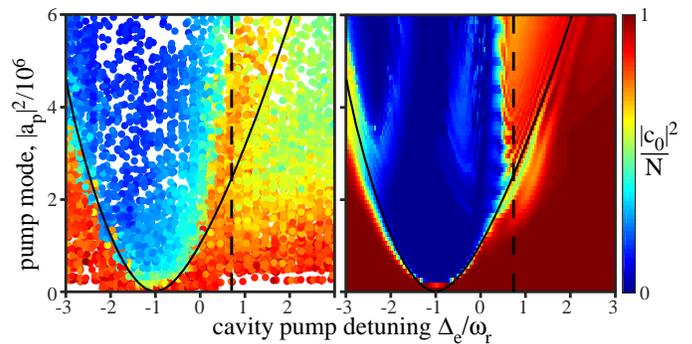}\caption{Phase
diagram. Population of the zero momentum state after 1.5ms of interaction with
the light in the resonator for various photon numbers in the pump mode and
cavity pump detuning. For negative detuning, the experimental observations
(left subplot) are well described by the phase boundary derived from a
numerical simulation that does not include the pinning potential (solid line).
The simulation shown in the right subplot includes the pinning potential. The
dashed line indicates the critical detuning within the limit of strong pumping
according to Eq. (\ref{kritischeVerstimmung}). }%
\label{Phase_diagram}%
\end{figure}Data are taken from 20\,000 experimental cycles for various cavity
pump detunings $\Delta_{c}$ and photon numbers $\left|  a_{p}\right|  ^{2}$ in
the pump mode. The data are post selected according to the value of the ratio
$R:=\kappa_{s}/(2U_{0}N)$ for the specific cycle. The denominator contains the
total number of atoms $N$ and the single photon light shift $U_{0}=g_{\text{eff}}%
^{2}/\Delta_{a}$ with the coupling constant $g_{\text{eff}}=\frac{\omega d^{2}%
}{6\hbar\varepsilon_{0}V}=2\pi\times$\SI{ 19}{\kilo\hertz}, the dipole moment
of the atomic transition $d$, and the permittivity of free space
$\varepsilon_{0}$. The ratio $R$ turns out to be the relevant parameter to
specify the strength of the pinning potential (see theory part below). Figure
\ref{Phase_diagram} shows the observed population $\left|  c_{0}\right|  ^{2}$
of the zero momentum state in the case of large mirror scattering ($R=0.15$),
a mean atom number in the BEC of $N=$\SI{1.8e5}{} and a atomic density of
\SI{5.7e12}{\centi\meter}$^{-3}$. The detuning $\Delta_{e}=\Delta_{c}+U_{0}N$
plotted along the horizontal axis is corrected for the index of refraction due
to the atoms. The blue area, where the system is unstable and almost all atoms
are excited into higher momentum states, is clearly separated from the stable
regime where at least half of the population $\left|  c_{0}\right|  ^{2}$
persists. For $\Delta_{e}=-$ $\omega_{r}$, the critical pump photon number for
entering the unstable regime is smallest since light scattered from the
initial condensate is recoil shifted by one $\omega_{r}$. For $\Delta
_{e}<-\omega_{r}$, the phase boundary between the stable and the unstable
regime follows the prediction of a numerical simulation, which ignores the
pinning potential (solid line in the left and right subplots). Evidently, the
pinning potential has only little effect in this regime. This is because at
threshold, the system jumps from a homogeneous superfluid state directly into
a state where the atoms form a density grating that moves with a finite start
velocity \cite{Dag14}. In the reference frame of the moving atoms, the pinning
potential averages out and has no effect. On the contrary, for positive
detuning, the atoms form a stationary density grating which can be seeded
efficiently by the pinning potential. In fact, the observed phase boundary
steeply increases in this regime and asymptotically approaches a vertical line
positioned at a critical detuning of $\Delta_{0}\simeq0.7\omega_{r}$ (dashed
line in the left and right subplot). A numerical simulation which includes
pinning, reproduces this behavior (right subplot).

The theoretical analysis of the experiment describes the light in the pump
mode and the probe mode by the field operators $\hat{A}_{p}=\hat{a}_{p}%
e^{ikz}$ and $\hat{A}=\hat{a}e^{-ikz}$. The atomic matter field $\hat{\psi
}=\sum\hat{c}_{n}e^{2inkz}$ is expanded into momentum eigenstates, separated
by $2\hbar k$, which is the momentum transferred to the atoms by scattering a
single photon from the pump mode into the probe mode. The atoms and the light
interact via the optical dipole potential $H_{\text{int}}=\hbar U_{0}\int\hat{\psi
}^{+}\hat{\psi}\left(  \hat{A}_{p}+\hat{A}\right)  \left(  \hat{A}_{p}%
^{+}+\hat{A}^{+}\right)  dV$. Mirror scattering couples the pump mode with the
probe mode and forms the pinning potential, $H_{p}=-\hbar\kappa_{s}\left(
\hat{A}_{p}\hat{A}+\hat{A}_{p}^{+}\hat{A}^{+}\right)  $. The equations of
motion are derived from the Hamiltonian $H=H_{0}+H_{int}$ $+H_{p}$ with
$H_{0}=\int\left[  \psi^{+}\left(  -\hbar^{2}\nabla^{2}/\left(  2m\right)
\right)  \psi+\hbar\Delta_{c}\left(  A^{+}A+A_{p}^{+}A_{p}\right)  \right]
dV$. Since the chemical potential of the condensate is much smaller than the
recoil energy, the small contributions due to atom-atom interaction are
neglected. In mean field approximation, operators are replaced by their
expectation values $a_{p}$, $a $, and $c_{n}$. Since the power of the pump
mode is electronically stabilized we set $a_{p}$ to be constant. Because only
the relative phase between $a$ and $a_{p}$ is physically relevant we also set
$a_{p}=\left|  a_{p}\right|  $. For the equations of motion one then gets
\cite{Piovella01}
\begin{subequations}
\begin{align}
\dot{c}_{n}  &  =-in^{2}\omega_{r}c_{n}-i\sigma\left(  c_{n-1}a^{\ast}%
+c_{n+1}a\right) \label{CARL-equations}\\
\dot{a}  &  =-\left(  \kappa+i\Delta_{e}\right)  a-i\sigma\sum_{n}c_{n}^{\ast
}c_{n-1}-i\kappa_{s}\left|  a_{p}\right| \label{CARL-equations_light}%
\end{align}
with the coupling constant $\sigma:=U_{0}\left|  a_{p}\right|  $ and the total
number of atoms $N=\sum c_{n}^{\ast}c_{n}$. The finite cavity line width is
taken into account by adding the decay term $-\kappa a$. The simulations in
Fig. \ref{Phase_diagram} are based on Eq. (\ref{CARL-equations},
\ref{CARL-equations_light}) with the sum ranging from $n=-5$ to $n=5$, since
higher momentum states have not been observed for the chosen experimental parameters.

To gain further physical insight we interpret Eq. (\ref{CARL-equations},
\ref{CARL-equations_light}) in the vicinity of the threshold. Higher momentum
states with $\left|  n\right|  >1$ can then be neglected yielding
\end{subequations}
\begin{subequations}
\begin{align}
\dot{c}_{-1}  &  =-i\omega_{r}c_{-1}-i\sigma ac_{0}%
\label{Viermodengleichungen}\\
\dot{c}_{0}  &  =-i\sigma\left(  ac_{1}+a^{\ast}c_{-1}\right) \\
\dot{c}_{1}  &  =-i\omega_{r}c_{1}-i\sigma a^{\ast}c_{0}\\
\dot{a}  &  =-i\Delta a-i\sigma b-i\kappa_{s}\left|  a_{p}\right|
.\label{Viermodengleichung_d}%
\end{align}
Here, we introduce the complex detuning $\Delta=\left|  \Delta\right|
e^{i\delta}:=\Delta_{e}-i\kappa$ and the complex structure factor $b=\left|
b\right|  e^{i\varphi_{b}}:=c_{1}^{\ast}c_{0}+c_{0}^{\ast}c_{-1}$. Without
mirror scattering, the population of the zero momentum component $\left|
c_{0}\right|  ^{2}$ remains undepleted until the system becomes unstable. In
previous work, $c_{0}$ was thus approximated as constant near threshold. The
equations then become linear and can be solved analytically \cite{Moore99, Dimer07}. 
If mirror scattering is included, the resulting optical
lattice potential depletes the zero momentum component even below threshold.
Thus $c_{0}$ has to be kept variable and the equations resume their nonlinear
character. Treating Eq. (\ref{Viermodengleichungen}-\ref{Viermodengleichung_d}%
) by linearization around the steady state solutions ($\dot{a}=\dot{c}_{0,\pm
1}=0$) is not successful since a constant structure factor may exist, even if
the coefficients $c_{\pm1},_{0}$ are time dependent. The stability diagram can
still be derived with the following strategy. In a first step, we solve the
first three equations with $a=\left|  a\right|  e^{i\varphi_{a}}$ being
regarded as a time independent parameter. The resulting linear eigenproblem
can then be solved straight forward. It has three eigenstates, with one of
them being a dark state that does not couple to the light field. The modulus
of the structure factor for the two other states can be calculated to be
\end{subequations}
\begin{equation}
B=\frac{1}{\sqrt{8}}\frac{A/A_{s}}{\sqrt{1+\left|  A\right|  ^{2}/A_{s}^{2}}%
}.\label{Strukturfaktor}%
\end{equation}
\begin{figure}[t]
\includegraphics[width=8cm]{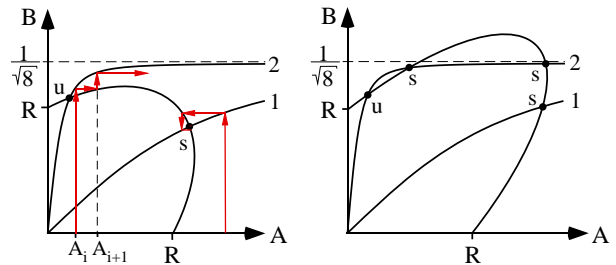}\caption{Relation
between the amplitudes of probe mode A and of structure factor B for weak and
strong pumping (saturation curve 1 and 2) aswell as small and large cavity
pump detuning (left and right subplot). The stability of the equilibrium
points at the intersection of the saturation curves with the ellipse (dots)
can be determined geometrically (red arrows). }%
\label{diagrams}%
\end{figure}Here, we introduce the normalized strength of the structure factor
$B:=\left|  b\right|  /(2N)$, the normalized amplitude of the light mode $A:=$
$\left|  a\right|  \left|  \Delta\right|  /(2N\sigma)$ and the saturation
parameter $A_{s}:=\left|  \Delta\right|  \omega_{r}/\left(  2\sqrt{8}%
N\sigma^{2}\right)  $. For both states the structure factor is time
independent and saturates at a maximum $B_{m}:=1/\sqrt{8}$ as $A_{s}$
approaches zero for strong pumping. The two states differ in the limit of
vanishing $a$, where the population $\left|  c_{0}\right|  ^{2}$ approaches
either zero or $N$. We thus ignore the first case since in the experiment all
atoms are initially in the condensate. The calculation shows that for the
second case the phases of the structure factor and the light field are equal,
$\varphi_{b}=\varphi_{a}$. In a second step we determine how, vice versa, a
given structure factor leads to a stable light field. Setting $\dot{a}=0$ in
Eq. (\ref{Viermodengleichung_d}) yields
\begin{equation}
A^{2}+B^{2}+2AB\cos\delta=R^{2}.\label{Ellipsengleichung}%
\end{equation}

In Fig. \ref{diagrams}, Eq. (\ref{Strukturfaktor}) and
(\ref{Ellipsengleichung}) are plotted. Eq. (\ref{Ellipsengleichung}) forms an
ellipse tilted by \SI{45}{\degree}. Its long axis varies between $2R$ (circle)
for $\Delta_{e}=0$ and infinity for $\Delta_{e}\gg\kappa$. Equilibrium states
exist at the intersection points of both curves. \begin{figure}[t]
\includegraphics[width=8cm]{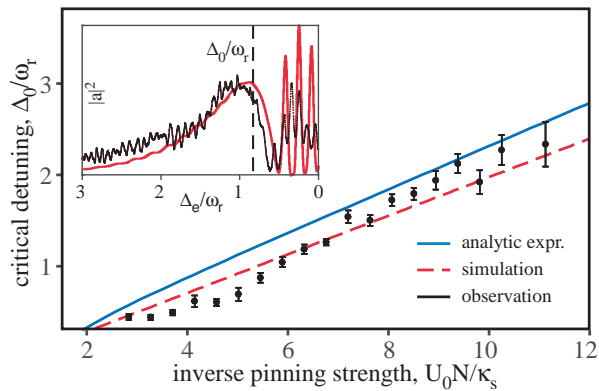}\caption{Shift of the
phase boundary $\Delta_{0}$ with the inverse strength of the pinning
potential. The observations (black dots), the simulations (red dashed line),
and the analytic expression of Eq. (\ref{kritischeVerstimmung}) are in
reasonable agreement. The inset shows the observed power in the probe mode
during the quench (black line). We identify the phase boundary $\Delta_{0}$
(dashed vertical line) as the position of the quick drop that follows the
continuous increase while $\Delta_{e}$ approaches $\Delta_{0}$ from above. The
rapid oscillations for $\Delta_{e}<\Delta_{0}$ are typical for the unstable
regime. The red line shows the result of a numerical simulation. }%
\label{Quench}%
\end{figure}The stability of equilibrium points are determined by reading from
the diagram how a given field $A_{i}$ results in a structure factor $B$
(vertical arrows) and how the so determined $B$ generates a new light field
$A_{i+1}$ (horizontal arrows). By repeating this sequence, the resulting
series $A_{i}$ converges for stable equilibrium and diverges otherwise. For
small detuning (left subplot) and weak pumping (saturation curve 1) one finds
a single point of stable equilibrium (indicated by ''s''). For stronger
pumping, the point moves to smaller $A$ and eventually becomes unstable
(saturation curve 2, ''u''). Without condensate depletion (neglecting the
second term in the square root of Eq. (\ref{Strukturfaktor})) the system
becomes unstable for $A_{s}=1/\sqrt{8}$ which reproduces the threshold
behavior found in previous models \cite{Moore99, Dimer07, Dag14}. For large detuning, the stable point remains stable even for
large pumping. This is true for arbitrary pump strength only if the maximum of
the ellipse $R/\sin\left(  \delta\right)  $, exceeds the maximum value of the
structure factor $B_{m}$ (dashed line). This condition determines the critical
detuning $\sin\left(  \delta_{0}\right)  =R/\sqrt{8}$. After replacing the
above definitions, the critical detuning defining the vertical phase boundary
in the limit of strong pumping reads
\begin{equation}
\frac{\Delta_{0}}{\kappa}=\sqrt{\frac{1}{8}\left(  \frac{2U_{0}N}{\kappa_{s}%
}\right)  ^{2}-1}\text{.}\label{kritischeVerstimmung}%
\end{equation}
It depends on the strength of the pinning potential $\kappa_{s}$ via the ratio
$1/R=2U_{0}N/\kappa_{s}$. Compared to this analytical expression, numerical
analysis shows a shift of the threshold to smaller detunings for lower pumping
strengths. We tested this relation by recording the phase boundary for various
mirror scattering $\kappa_{s}$ and atom number $N$. The phase boundary is
detected by sweeping the detuning $\Delta_{e}$ from large to small values
within \SI{1}{\milli\second}, while the photon number in the pump mode is
electronically stabilized to a constant value of $\left|  a_{p}\right|  ^{2}%
=$\SI{4e6} (inset in Fig. \ref{Quench}). While sweeping, the power in the
probe mode increases until eventually the threshold is reached. We identify
the critical detuning $\Delta_{0}$ at the edge, where the power in the probe
mode drops quickly and the system becomes unstable. By repeating the
experiment for various values of $NU_{0}/\kappa_{s}$ we obtain the curve in
the main graph of Fig. \ref{Quench}. The observations (black dots), the
simulation (red dashed line) and the analytic expression (Eq.
(\ref{kritischeVerstimmung}), blue line) are in reasonable agreement. In the
absence of a pinning potential, previous theoretical models \cite{Moore99}
predict strict threshold behavior only for lossless cavities. If losses are
included, the threshold smears out and the system becomes unstable even for
infinitesimally low pump power. Our observations, however, support a physical
picture (Fig. \ref{diagrams}) that predicts strict threshold behavior also for
lossy cavities and even in the limit of vanishing injection ($R \rightarrow0$).

In summary, we have investigated an atomic Bose Einstein condensate in an
optical ring resonator with additional pinning potential. A stable phase was
identified above a critical cavity pump detuning. The phase boundary is
defined by the competition of the pinning potential and the optical potential
generated by the atoms. The observations are quantitatively described by
simulating the nonlinear equations of motion, including depletion of the
condensate. A geometric interpretation is introduced to determine equilibrium
and stability of the system and an analytic expression for the phase boundary
is derived in the limit of strong pumping. By seeding the probe mode, the
transition from a ring geometry to a standing wave geometry can be explored
similar as in recent work with a condensate replaced by a nano membrane
\cite{Yilmaz16}. More work is required to understand the role of the two
additional points of equilibrium which appear above the critical detuning.
Also unclear is the classification of the phase transition, quantum
fluctuations near threshold and possible metastability \cite{Hruby17}.

We acknowledge helpful discussion with Andreas Hemmerich, Martin Schmidt and
Peter Domokos. This work has been supported by the Deutsche
Forschungsgemeinschaft (ZI 419/8-1) and by ColOpt -- EU H2020 ITN 721465.

%\newpage

\end{document}